\documentclass[aps,prb,twocolumn,footinbib,superscriptaddress]{revtex4}
\usepackage{graphicx}
\usepackage{spreadtab}       
\usepackage{dsfont}
\usepackage{color}           
\usepackage{hyperref}        
\hypersetup{
    colorlinks=true,
    linkcolor=blue,
    citecolor=blue,
    urlcolor=blue,
}
\usepackage{amsmath}         
\usepackage{booktabs}        

\newcommand\bk{{\mathbf{k}}}
\newcommand\bq{{\mathbf{q}}}

\newcommand\bR{{\mathbf{R}}}

\begin{document}

\title{Impact of anharmonicity on the carrier mobility of the Pb-free CsSnBr$_3$ perovskite}

\author{Junwen Yin}
\affiliation{%
European Theoretical Spectroscopy Facility, Institute of Condensed Matter and Nanosciences, Université catholique de Louvain, Chemin des Étoiles 8, B-1348 Louvain-la-Neuve, Belgium. 	
}%
\author{Olle Hellman}
\affiliation{%
Department of Molecular Chemistry and Materials Science, Weizmann Institute of Science, Rehovot 7610001, Israel.		
}%
\author{Samuel Ponc\'e}
\email{samuel.ponce@uclouvain.be}
\affiliation{%
European Theoretical Spectroscopy Facility, Institute of Condensed Matter and Nanosciences, Université catholique de Louvain, Chemin des Étoiles 8, B-1348 Louvain-la-Neuve, Belgium. 	
}%
\affiliation{%
WEL Research Institute, avenue Pasteur 6, 1300 Wavre, Belgique.			
}%

\date{\today}

\begin{abstract}

Charge carrier mobilities are critical parameters in halide perovskite solar cells, governing their average carrier velocity under an applied electric field and overall efficiency. 
Recent advances in first-principles calculations of electron-phonon interactions and carrier mobilities have enabled predictive computations for perovskite solar cells. 
However, the flexible octahedral frameworks and cationic displacements in these materials challenge the harmonic approximation, leading to significant difficulties in accurately calculating transport properties. 
To address these issues, we combine temperature-dependent effective potentials with the \textit{ab initio} Boltzmann transport equations to compute carrier mobilities in a representative lead-free perovskite, CsSnBr$_3$. 
At room temperature, the electron/hole Hall mobilities in CsSnBr$_3$ are 106/256~cm$^2$/Vs when neglecting anharmonic effects and 59/145~cm$^2$/Vs when included. 
This overestimation of the harmonic approximation arises from the neglect of scattering coming from soft modes.
We provide a workflow for performing first-principles carrier mobility calculations in anharmonic systems, advancing the predictive modeling of perovskite solar cells.
\end{abstract}

\maketitle

\section{Introduction}

Halide perovskites have emerged as a highly promising class of materials, particularly in the field of optoelectronics~\cite{Stranks2015, Rong2018}. 
These materials have attracted significant attention for their remarkable properties and potential applications, especially in solar cells, light-emitting diodes, and photodetectors~\cite{Nie2015}. 
Halide perovskites have several exceptional properties, including a high absorption coefficient, long carrier diffusion lengths, tunable bandgap, and solution processability~\cite{Kumar2024}. 
The most popular perovskites today are lead-based, particularly methylammonium and formamidinum lead halides ([MA,FA]PbX$_3$), known for their outstanding optoelectronic properties~\cite{Torrence2023,Ren2022}. 
However, the toxicity of lead poses serious environmental and health concerns~\cite{Torrence2023,Ren2022}, leading to a growing interest in lead-free alternatives, such as tin-, germanium-, and bismuth-based perovskites, to achieve similar performance while reducing risks~\cite{Liu2018}.
For lead-free perovskite solar cells, carrier mobility and recombination rates play a crucial role in determining the carrier diffusion length and the overall efficiency. 
High carrier mobility facilitates efficient charge transport, reducing losses and improving the performance of solar cells. 
However, compared to traditional inorganic semiconductors such as GaAs, which boast electron mobilities of up to 8000~cm$^2$/Vs and hole mobilities around 400~cm$^2$/Vs~\cite{Brenner2015}, lead-free perovskites typically exhibit much lower carrier mobilities, often in the range of tens of cm$^2$/Vs~\cite{Herz2017}. 
These relatively low mobilities arise from several factors, including defects, ion migration, and scattering at grain boundaries~\cite{Herz2017,Brenner2015,Ponce2019,Xia2021}. 
Understanding the intrinsic and extrinsic factors that limit carrier mobility in lead-free perovskites is essential to advance the design and fabrication of high-efficiency solar cells. 
Research efforts now focus on optimizing material composition, reducing defects, and improving intrinsic properties to enhance carrier mobility and minimize recombination losses~\cite{Tao2021,Wang2018}. 
Progress in this area is expected to boost the efficiency of lead-free perovskite solar cells and bring them closer to commercial viability.

Materials informatics has advanced rapidly in recent years, significantly aiding in the design and discovery of new materials. 
Techniques such as density functional perturbation theory (DFPT)~\cite{Gonze1997,Baroni2001}, Quantum Monte Carlo simulations~\cite{Foulkes2001}, and \textit{ab-initio} molecular dynamic (aiMD)~\cite{Iftimie2005} have been extensively employed to investigate the relationships between structural, compositional, and technological descriptors and material performance across various scales. 
However, in perovskite materials, the pronounced anharmonic lattice dynamics, driven by their inherently soft structures, pose significant challenges in accurately capturing electron-phonon interactions and carrier mobilities.
Several \textit{ab-initio} methods and software have been developed to address this issue, including self-consistent phonon (SCP)~\cite{Errea2014,Tadano2018} and self-consistent-field non-perturbative methods with temperature-dependent atomic displacement~\cite{Monserrat2013,Zacharias2023}, perturbative methods based on density functional perturbation theory to compute third-order force constants~\cite{Lazzeri2002,Paulatto2013}, and aiMD based methods~\cite{Hellman2011}.
There exist various flavors and software implementations of the SCP methods in which an effective harmonic force constant is obtained self-consistently by computing atomic forces in supercells.

These forces can be sampled using the stochastic self-consistent harmonic approximation (SSCHA~\cite{Errea2014}, \textsc{SSCHA} code~\cite{Monacelli2021})
or the temperature-dependent effective potential (TDEP~\cite{Hellman2011,Hellman2013}, \textsc{TDEP} code~\cite{Knoop2024}), 
both of which employ a stochastic framework for self-consistently determining anharmonic force constants. 
Alternatively, anharmonic forces can be obtained by Taylor expansion to the third order, which is applied in \textsc{phonopy}~\cite{Togo2015,Togo2023} and \textsc{D3Q}~\cite{Broido2007}. 
Another approach is the least absolute shrinkage and selection operator (LASSO, also known as compressed sensing)~\cite{Zhou2014,Tadano2015,Tadano2014,Masuki2023}, 
which provides an efficient framework to extracting higher-order interatomic force constants. 
The self-consistent \textit{ab-initio} lattice dynamics (SCAILD) method~\cite{Souvatzis2008} 
iteratively solves the self-consistent phonon (SCP) equations to determine the phonon properties in anharmonic systems. 
These methods have been implemented in various software packages such as \textsc{HiPhive}~\cite{Eriksson2019}, \textsc{ZG}~\cite{Zacharias2020}, and \textsc{a-TDEP}~\cite{Bottin2020}.

However, few of these anharmonic methods have been used to compute the charged carrier transport properties. 
The two most common \textit{ab-initio} methods to compute transport properties are the Green-Kubo formula~\cite{Green1954,Kubo1957,Lihm2025} or the iterative Boltzmann transport equation (BTE)~\cite{Ponce2020,Claes2025}.
In particular, the electron-phonon Wannier (\textsc{EPW}) code~\cite{Ponce2016,Lee2023} can efficiently interpolate Hamiltonian, dynamical matrices and electron-phonon matrix elements on dense momentum grids to solve the BTE.  
The Kubo-Greenwood approximation to Green-Kubo was recently used to calculate the carrier mobility of SrTiO$_3$ and BaTiO$_3$ using aiMD in the NVT canonical ensemble~\cite{Quan2024} and a charged carrier mobility expression based on MD mean-square displacement was also used to study MAPbI$_3$ and MAPbBr$_3$~\cite{Schilcher2023}.
Furthermore, using the BTE, the electron drift mobility of cubic SrTiO$_3$ was studied with anharmonic effects~\cite{Zhou2018}, while the electron and hole drift mobilities in CsSnI$_3$ and CsPbI$_3$ were studied using compressed sensing~\cite{Zhang2022}.

Building on these efforts, we have developed an interface between the \textsc{TDEP} and \textsc{EPW} software to include \emph{partial} anharmonic effects in carrier transport calculations. 
Here, by \emph{partial}, we mean that we use an updated effective interatomic force constant that includes anharmonic effects. 
Still, we compute the change of the perturbed potential using density functional perturbation theory (DFPT)~\cite{Gonze1997}, in line with current state-of-the-art approaches~\cite{Zhou2018}.  
In this work, we present calculations of the temperature-dependent phonon dispersion, electron-phonon interactions, and carrier mobilities of CsSnBr$_3$. 
This inorganic perovskite was selected as a starting point to explore lead-free perovskites due to its promising optoelectronic properties, including a suitable bandgap and the absence of toxic lead~\cite{Zheng1999,Coduri2019}. 
As one of the few tin-based halide perovskites, CsSnBr$_3$ shows potential as an environmentally friendly alternative while offering electronic structures comparable to those of its lead-based counterparts~\cite{Mori1986}.
However, oxidation and reduced stability have limited its broader application, making it an ideal candidate for further investigation to improve carrier mobility and stability through anharmonic lattice dynamics~\cite{Zhang2022}. 
We used DFPT and TDEP to obtain harmonic and anharmonic interatomic force constants (IFCs). 
The real-space effective IFCs are mapped into reciprocal space in our interface and subsequently used to calculate electron-phonon coupling matrix elements. 
We compute the mobility of CsSnBr$_3$ as a function of temperature and assess the impact of anharmonicity. 
We find that anharmonicity reduces the intrinsic Hall mobility and computes an electron and hole room-temperature mobility of 59~cm$^2$/Vs and 145~cm$^2$/Vs, respectively.

\section{Computational Methods}

Our investigation of the electronic structure and transport properties of halide perovskites was conducted using first-principles calculations within the framework of density functional theory (DFT) using the \textsc{Quantum ESPRESSO} (QE) package~\cite{Giannozzi2017}. 
We use fully relativistic optimized norm-conserving Vanderbilt pseudopotentials~\cite{Hamann2013} based on the Perdew-Burke-Ernzerhof (PBE) generalized gradient approximation~\cite{Perdew1996} from the standard \textsc{PseudoDojo} library~\cite{vanSetten2017}.
The plane-wave basis set was expanded with a kinetic energy cutoff of 80~Ry, and a $12 \times 12 \times 12$ \textbf{k}-point grid was used, ensuring convergence in energy with a tolerance of $10^{-11}$ Ry. 
The lattice parameters are optimized with PBE, including spin-orbit coupling (SOC), which gives a lattice parameter of 5.89~\text{\AA} for the cubic phase.
For the electronic bandstructure, we used the HSE06 hybrid functional~\cite{Heyd2003} with a 0.17 fraction of Fock exchange~\cite{Zhang2022} and a $6 \times 6 \times 6$ \textbf{k}-point grid. 
Harmonic phonon calculations are obtained using DFPT~\cite{Gonze1997} with a $6 \times 6 \times 6$ $\bq$-point grid. 
The dielectric functions and Born effective charge tensor computed at $\bq = \boldsymbol{\Gamma}$ require a dense $24 \times 24 \times 24$ \textbf{k}-point grid to be converged, see Fig.~S1 of the SI~\cite{Yin2025}.

To include anharmonic effects in the computations of phonon and electron-phonon coupling, we follow the TDEP method~\cite{Hellman2011}. 
The total energy can be written using a Taylor expansion around the equilibrium Born-Oppenheimer (BO) atomic position $\{\bR_0 \}$
\begin{multline}\label{eq:dm}
E^{\rm BO}_0\{\bR\} = E^{\rm BO}_0\{\bR_0\} \\
+ \frac{1}{2} \sum_{\substack{\kappa\alpha \\ \kappa'\alpha'}} \frac{\partial^2 E^{\rm BO}_0\{\bR\}}{\partial R_{\kappa\alpha}\partial R_{\kappa'\alpha'}}\bigg|_{\mathbf{0}} \Delta R_{\kappa\alpha}\Delta R_{\kappa'\alpha'} + \mathcal{O}( \Delta \bR^3),
\end{multline}
where $\Delta R_{\kappa\alpha} = R_{\kappa\alpha} - R_{0\kappa\alpha} $ is the displacement of atom $\kappa$ in the Cartesian direction $\alpha$, and where $|_{\mathbf{0}}$ indicates that we are taking the derivatives at $\{ \bR=\bR_0 \}$. 
The second-order derivative with respect to atomic displacement is the interatomic force constant (IFC), denoted by
$C_{\kappa\alpha\kappa'\alpha'}$.
The IFC can be determined using linear response methods or finite-difference, while the anharmonic higher-order terms are truncated. 

The TDEP approach consists of finding an effective harmonic IFC that best describes the energy landscape felt by the nuclei oscillating around their equilibrium position at a given temperature. 
To achieve this, a set of perturbed atomic configurations from equilibrium positions is generated for each temperature and volume using stochastic TDEP (sTDEP)~\cite{Shulumba2017}. 
To determine the effective force constant matrix \(\tilde{C}\) that most accurately represents the real forces, we minimize the force difference over $N$ configurations
$\sum_i^N \frac{1}{N}\sum_{\kappa} | \mathbf{F}_\kappa - \tilde{\mathbf{F}}_\kappa |$, where $\mathbf{F}_\kappa$ is the DFT force on atom $\kappa$ and $\tilde{\mathbf{F}}_\kappa $ is the effective force computed
from 
\begin{equation}
\tilde{F}_{\kappa\alpha}  = \sum_{\kappa'\alpha'} \tilde{C}_{\kappa\alpha,\kappa'\alpha'} \Delta R_{\kappa'\alpha'}.
\end{equation}
If the number of configurations exceeds three times the number of atoms, the system of equations for $\tilde{C}$ becomes overdetermined. 
We obtain the linear least-squares solution for $\tilde{C}$. 
After convergence with respect to the number of configurations, we define the force constant matrix at temperature $T$
and use it to generate new stochastic displacements. 

In this work, we study CsSnBr$_3$ in its cubic phase, which is stable above 292~K~\cite{Mori1986}.
Therefore, we performed sTDEP calculations at five different temperatures: 300~K, 350~K, 400~K, 450~K, and 500~K.
We first compute the force constants at 500~K since large displacements are easier to converge and use them as initial guesses for lower-temperature calculations. 
Seven iterations are required for each temperature to converge the phonon dispersions to 0.1~meV.
Each iteration uses $N=2^l + 2^{(l-1)}$ configurations, where $l$ is the iteration number.

When unbound electrons are subject to an external electric field in semiconductors at finite temperatures, they move and experience scattering due to interactions with lattice vibrations, also called phonons. 
This scattering disrupts their motion, contributing to resistivity and defining the carrier mobility. 
We use the BTE as implemented in the \textsc{EPW} code to study the low-field mobility of semiconductors, where the drift mobility $\mu_{\alpha \beta}$ is obtained from the electrical conductivity tensor $\sigma_{\alpha \beta}$ divided by the carrier density \(n_c\): $\mu_{\alpha \beta} = \frac{\sigma_{\alpha \beta}}{e n_c}$.
In this equation, the electrical conductivity tensor $\sigma_{\alpha \beta}$ corresponds to the variation of the carrier occupation function $f_{n\bk}$ with respect to the electric field $\mathbf{E}$, being computed using~\cite{Ponce2018}:
\begin{equation}\label{eq:conductivity}
\sigma_{\alpha \beta} = - \frac{e}{\Omega^{\rm uc}} \sum_n \int \frac{\rm d \mathbf{k}}{\Omega^{\rm BZ}} v_{n\mathbf{k}\alpha} \partial_{E_\beta} f_{n\mathbf{k}}
\end{equation}
where $\Omega^{\rm BZ}$ is the Brillouin-Zone volume, $\Omega^{\rm uc}$ the unit cell volume, $\partial_{E_\beta} f_{n\mathbf{k}}$ is the linear response of the carrier occupation function due to an external electric $\mathbf{E}$, given by~\cite{Ponce2018,Ponce2021,Lee2023}:
\begin{multline}\label{eq:iterative}
\partial_{E_\beta} f_{n\mathbf{k}} = e v_{n\mathbf{k}\beta} \frac{\partial f^0_{n\mathbf{k}}}{\partial \epsilon_{n\mathbf{k}}} \tau_{n\mathbf{k}} 
+ \frac{2\pi \tau_{n\mathbf{k}}}{\hbar} \! \sum_{m\nu} \int \! \frac{d\mathbf{q}}{\Omega^{\rm BZ}} |g_{mn\nu}(\mathbf{k}, \mathbf{q})|^2 \\
\times \left[ \left(n_{\bq\nu} + 1 - f^0_{n\mathbf{k}} \right) \delta(\epsilon_{n\mathbf{k}} - \epsilon_{m\mathbf{k+q}} + \hbar \omega_{\bq\nu}) \right. \\
+ \left. \left(n_{\bq\nu} + f^0_{n\mathbf{k}} \right) \delta(\epsilon_{n\mathbf{k}} - \epsilon_{m\mathbf{k+q}} - \hbar \omega_{\bq\nu}) \right]\partial_{E_\beta} f_{m\mathbf{k+q}},  
\end{multline}
where $n$ denotes a Kohn-Sham band index, $\mathbf{k}$ is the electron wavevector, $\partial_{E_\beta} f_{n\mathbf{k}}$ is a short-hand notation for $(\partial f_{n\mathbf{k}} / \partial E_\beta)|_{E=0}$,
$v_{n\mathbf{k}\alpha} = \hbar^{-1} \partial \epsilon_{n\mathbf{k}} / \partial k_\alpha$ is the intra-band velocity matrix element for the Kohn-Sham eigenvalue $\epsilon_{n\mathbf{k}}$, and $\delta$ denotes the Dirac delta function, $g_{mn\nu}(\mathbf{k}, \mathbf{q})$ is the electron-phonon matrix element describing the interaction from the state $n\mathbf{k}$ to the state $m\mathbf{k} + \mathbf{q}$ via the phonon of branch $\nu$ and wavevector $\mathbf{q}$. 
The temperature enters this equation via the Fermi-Dirac and Bose-Einstein equilibrium distribution functions $f^0_{n\mathbf{k}}$ and $n_{\bq\nu}$, respectively. 
The quantity $\tau_{n\mathbf{k}}$ in Eq.~\eqref{eq:iterative} is the carrier relaxation time and is obtained from Fermi’s golden rule:
\begin{multline}\label{eq:scat}
\tau_{n\mathbf{k}}^{-1} = \frac{2\pi}{\hbar} \sum_{m\nu} \int \frac{d\mathbf{q}}{\Omega^{\rm BZ}} |g_{mn\nu}(\mathbf{k}, \mathbf{q})|^2 \\
 \times \left[ \left( n_{\bq\nu} + 1 - f^0_{m\mathbf{k+q}} \right) \delta(\epsilon_{n\mathbf{k}} - \epsilon_{m\mathbf{k+q}} - \hbar \omega_{\bq\nu}) \right. \\
 + \left. \left( n_{\bq\nu} + f^0_{m\mathbf{k+q}} \right) \delta(\epsilon_{n\mathbf{k}} - \epsilon_{m\mathbf{k+q}} + \hbar \omega_{\bq\nu}) \right].
\end{multline}
However, we note that often it is the Hall and not the drift mobility which is measured in experiment.
To compute the Hall mobility, we add a vanishing finite magnetic field in Eq.~\eqref{eq:iterative} following Refs.~\onlinecite{Ponce2021,Lee2023}.
In this work, we interpolated the electron-phonon matrix element to compute the drift and Hall mobility using 80 $\times$ 80 $\times$ 80 fine $\bk$ and $\bq$ grids to reach convergence, shown in Fig.~S2 of the SI~\cite{Yin2025} for the case of hole mobility.

\section{Results and Discussion}

The CsSnBr$_3$ perovskite undergoes multiple phase transitions, evolving from monoclinic at low temperature to tetragonal at 247~K and from tetragonal to cubic (Pm$\bar{3}$m, No. 221) at 292~K~\cite{Mori1986}.
Large Cs$^+$ ions occupy the voids within the corner-sharing SnBr$_6$ octahedral framework in the cubic phase. 
Anharmonicity in this phase arises from significant atomic displacements and dynamic tilting of the SnBr$_6$ octahedra, leading to deviations from the ideal harmonic behavior, which influence the thermal and vibrational properties of the material.
As inferred from Eqs.~\eqref{eq:conductivity}-\eqref{eq:scat}, accounting for the effects of anharmonicity in electron-phonon interactions and carrier mobility requires the accurate calculation of the electron-phonon matrix elements $g_{mn\nu}(\mathbf{k}, \mathbf{q})$, which are expressed as~\cite{Giustino2017}:
\begin{equation}
\! g_{mn\nu}(\mathbf{k}, \mathbf{q}) = \sqrt{\frac{\hbar}{2 \omega_{\mathbf{q} \nu}}} \! \sum_{\kappa \alpha} \frac{e_{\kappa\alpha,\nu\mathbf{q}}}{\sqrt{M_{\kappa}}} \langle \Psi_{m\mathbf{k} + \mathbf{q}} | V_{\mathbf{q} \kappa \alpha} | \Psi_{n \mathbf{k}} \rangle,
\end{equation}
with $V_{\mathbf{q} \kappa \alpha}$ the variation of the Kohn-Sham potential due to displacements of atom $\kappa$ (with mass $M_{\kappa}$) in the Cartesian direction $\alpha$. 
In this work, we use TDEP to compute anharmonic phonon energies $\omega_{\mathbf{q} \nu}$ and eigenvectors $e_{\kappa\alpha,\mathbf{q} \nu}$.
However, we do not update the deformation potential $V_{\mathbf{q} \kappa \alpha}$ and keep it at the harmonic level. 
Due to the presence of heavy elements, we perform calculations with and without spin-orbit coupling (SOC) for comparison.
Since it is known that the PBE functional significantly underestimates the band gap, we also perform hybrid functional calculations using HSE06~\cite{Heyd2003}.
The phonon properties are subsequently calculated using both DFPT and TDEP to highlight the impact of anharmonicity. 
Electron-phonon interactions are interpolated using \textsc{EPW}. 
To ensure the accuracy of these interpolations, we benchmarked the electron-phonon coupling matrix elements against the DFPT results. 
Finally, carrier transport properties are investigated using BTE.

\subsection{Electronic Band Structures and phonon dispersion}
\begin{figure}[t] 
  \centering 
  \includegraphics[width=0.9\linewidth]{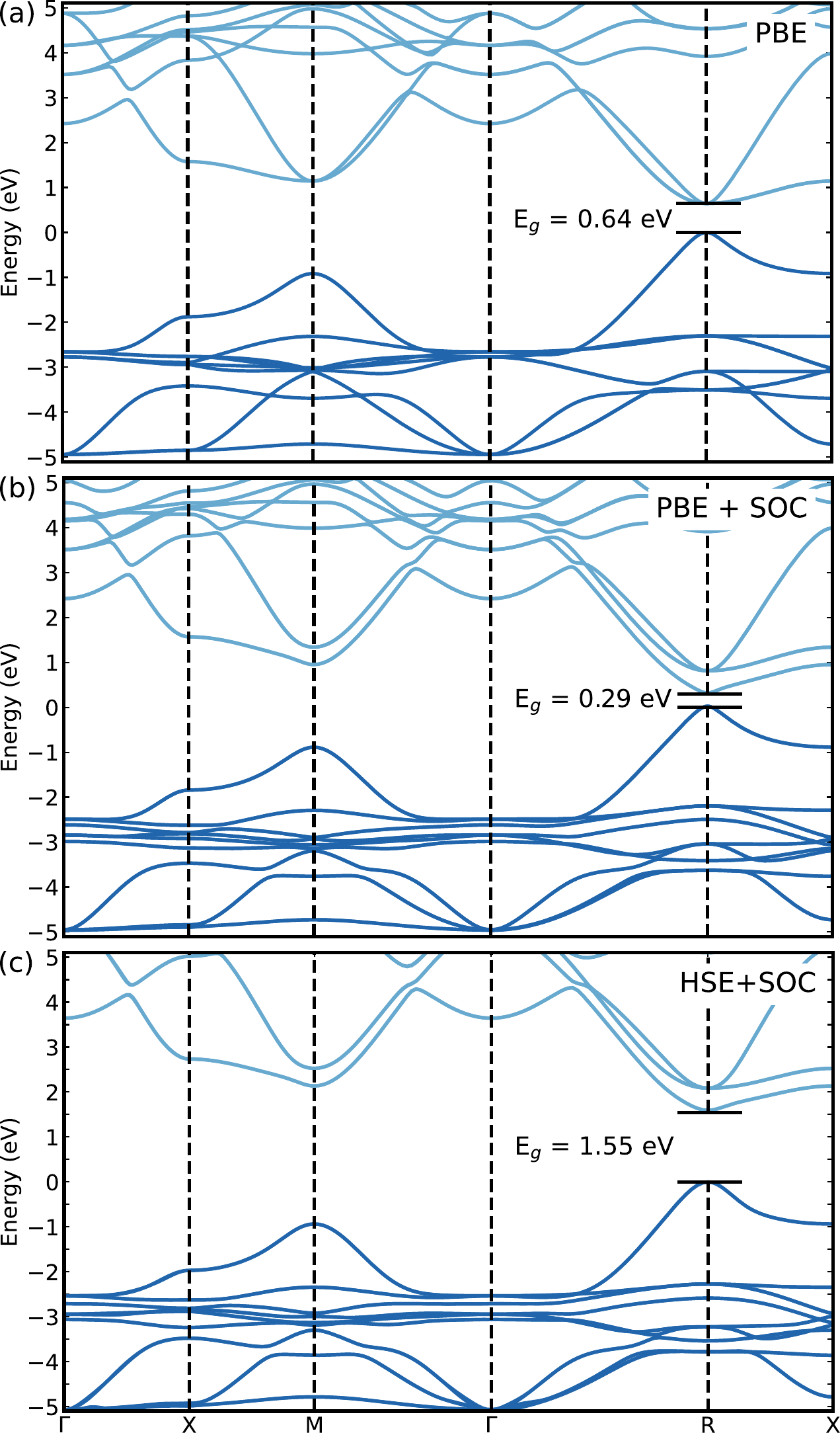} 
  \caption{\label{fig1}
  Electronic band structures of cubic CsSnBr$_3$ (a) calculated with PBE without SOC, (b) with PBE with SOC, and (c) with HSE06 with SOC. 
  The valence band maximum is set to zero as the reference energy level.
  } 
\end{figure}

We obtained a lattice constant of 5.89~\text{\AA} from PBE calculations, both with and without SOC, while measurements report a lattice constant of 5.80~\text{\AA}~\cite{Zheng1999}. 
As shown in Fig.~\ref{fig1}, cubic CsSnBr$_3$ exhibits a direct band gap located at the $R$ point of the Brillouin zone. 
The band gap (E$_g$) is calculated to be 0.64~eV at the PBE level without SOC, reducing to 0.29~eV when SOC is included. 
Both values are significantly lower than the experimental value of 1.36~eV~\cite{Coduri2019}. 
At the HSE06 level with SOC included, the band gap increases to 1.55~eV, in better agreement with experimental reports.

The band edge character is dominated by the Sn-Br bonds, with minimal contribution from Cs. 
The edge of the conduction band originates from the anti-bonding states between the 6s orbitals of Br and the 5p orbitals of Sn. 
In comparison, the edge of the valence band is mainly contributed by the anti-bonding states between the 5p orbitals of Br and the 5s orbitals of Sn. 
Without SOC, the CBM is threefold degenerate. 
When SOC is introduced, it splits into a doublet and a quadruplet. 
As the band edge is crucial for carrier mobility calculations, we include SOC in electron-phonon interaction and carrier transport investigations.  

\begin{figure}[t] 
  \centering 
  \includegraphics[width=0.9\linewidth]{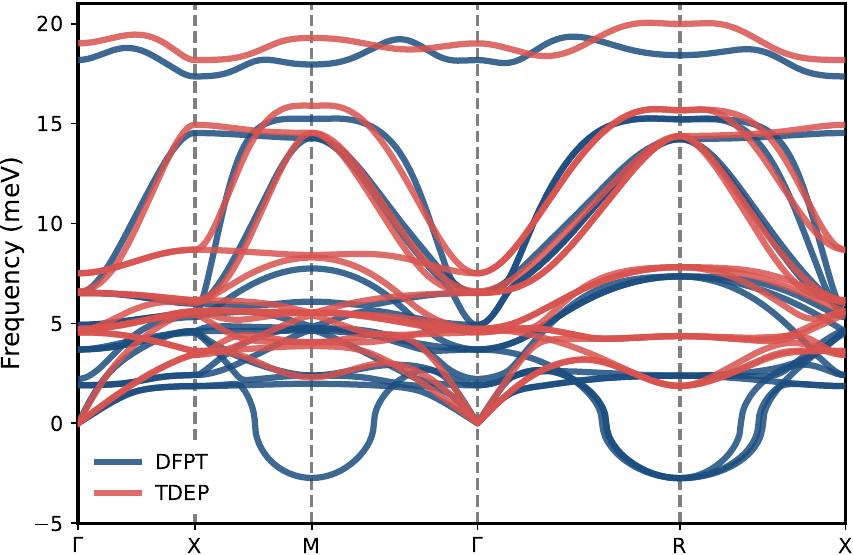} 
  \caption{\label{fig2} 
  Phonon band structures of cubic CsSnBr$_3$ where the blue lines show the harmonic results while the red lines show the anharmonic results at 300~K.
}   
\end{figure}

\begin{figure}[t] 
  \centering 
  \includegraphics[width=0.9\linewidth]{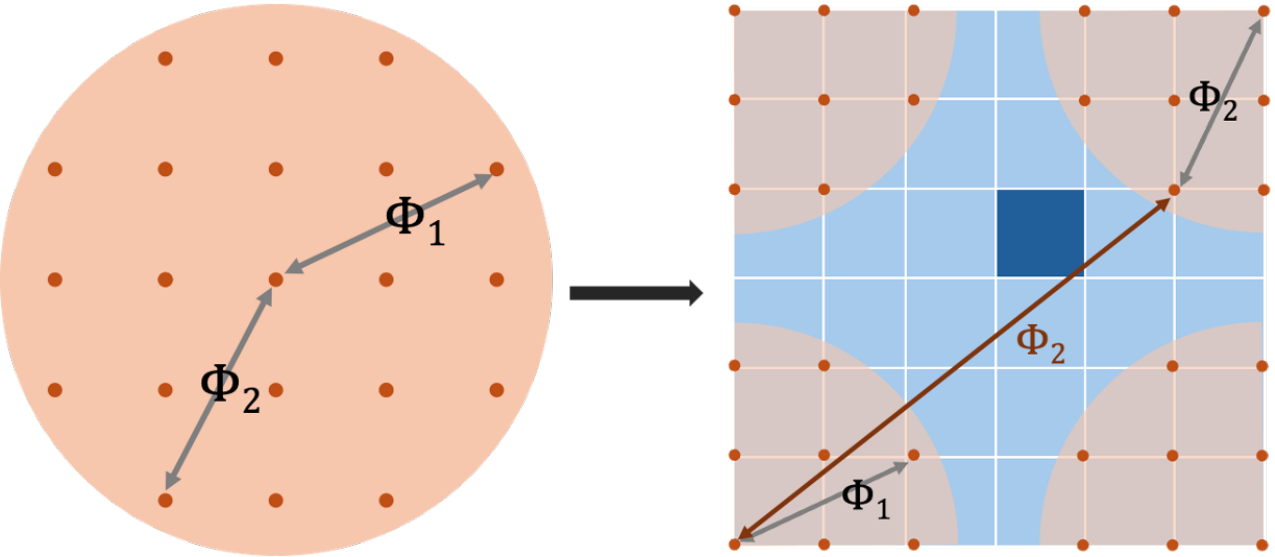} 
  \caption{\label{fig3} Two dimensional illustration of TDEP force constant ($\Phi$) conversion into \textsc{Quantum ESPRESSO} format. 
  The left figure represents the TDEP supercell in real space, where the orange dots depict atoms, and the circle's radius indicates the real space cutoff for the force constants. 
  The right figure shows the \textsc{Quantum ESPRESSO} supercell in reciprocal space where the deep blue square represents the unit cell. 
} 
\end{figure}

Figure ~\ref{fig2} compares the phonon dispersions calculated using DFPT and TDEP at 300~K, and we show the results for the other temperatures in the SI~Fig.~S3~\cite{Yin2025} where we observe a stiffening of the low-frequency modes with temperature.
Both calculations account for the LO-TO splitting. 
Consistent with previous reports on perovskite structures~\cite{Zhang2022, Zhou2018}, the harmonic results show negative frequencies at the $R$ and $M$ points, indicating cation rattling and octahedral instability. 
In contrast, the results which includes anharmonicity at 300~K eliminate these negative frequencies, stabilizing the soft phonons.
Notably, the lowest-energy optical modes at the zone center more than double in frequency, while the highest-energy optical modes remain almost constant upon inclusion of anharmonicity. 
To integrate the TDEP IFCs into \textsc{Quantum ESPRESSO} and \textsc{EPW} for electron-phonon interaction calculations, 
we map the spherical IFCs from \textsc{EPW}  onto the edge-centered convention used by \textsc{Quantum ESPRESSO} as shown in Fig.~\ref{fig3}.
Dipolar long-range interactions are removed and consistently re-added during Fourier interpolation following the method of Refs.~\onlinecite{Gonze1994,Gonze1997} with identical Ewald parameters set in both codes.

\subsection{Effective Mass and  Wannier Interpolation}

\begin{table}[t]
  \caption{\label{table1} Effective masses of cubic CsSnBr$_3$ in units of the electron rest mass along the (111) direction, corresponding to the $R-\Gamma$ direction and the (001) direction corresponding to the $R-X$ direction.
  }
  \centering
  \setlength{\tabcolsep}{5pt} 
  \renewcommand{\arraystretch}{1.2} 
  \begin{tabular}{cccccccc}
    \hline
    Direction & \multicolumn{3}{c}{PBE} & \multicolumn{2}{c}{PBE+SOC} & \multicolumn{2}{c}{HSE+SOC} \\ 
              & $m^{\ast}_{h}$ & $m^{\ast}_{le}$ & $m^{\ast}_{he}$ & $m^{\ast}_{h}$ & $m^{\ast}_{e}$ & $m^{\ast}_{h}$ & $m^{\ast}_{e}$ \\
    \hline
    (111)     & 0.07           & 0.08            & 0.27            & 0.06           & 0.09           & 0.13           & 0.27 \\
    $(0\,\bar{1}\,\bar{1})$      & 0.07           & 0.05            & 0.84            & 0.06           & 0.08           & 0.13           & 0.22 \\
    \hline
  \end{tabular}
\end{table}

We fit the band edges to a parabolic function and report the effective mass of carriers in CsSnBr$_3$ \(m^{\ast}_{h(e)}\) in Table~\ref{table1}. 
We compare the calculated effective masses for the CsSnBr$_3$ system under three conditions: PBE without SOC, PBE with SOC, and HSE06 with SOC. 
Specifically, we focus on holes at the valence band maximum (VBM) and electrons at the conduction band minimum (CBM) at the $R$ point in the Brillouin zone. 
For the VBM, the band dispersion is isotropic and parabolic across all scenarios. 
In particular, the effective masses of holes, $m^{\ast}_{h}$, range from 0.06 to 0.13, consistently lower than the corresponding electron effective masses, $m^{\ast}_{e}$, in all cases. 
Given the Drude mobility $\mu = \frac{e \tau}{m^*}$, this suggests a higher hole mobility compared to that of electrons, a prediction that will be corroborated by the transport calculations discussed later.

The discussion of the electron effective mass, $m^{\ast}_{e}$, reveals a more complex situation than for $m^{\ast}_{h}$. 
In the absence of SOC, the CBM is threefold degenerate, resulting in distinct heavy electron (\(m^{\ast}_{he}\)) and light electron ($m^{\ast}_{le}$) masses.
Furthermore, $m^{\ast}_{e}$ exhibits anisotropy depending on the measurement direction. 
For $m^{\ast}_{le}$, the effective masses are similar along the (111) and (001) directions, with values of 0.08 and 0.05, respectively. 
However, for $m^{\ast}_{he}$, a significant anisotropy is observed, with the effective mass increasing from 0.27 along the (111) direction to 0.84 along the (001) direction, nearly tripling in magnitude. 
Based on the calculated results for the effective mass of CsSnBr$_3$, we infer that holes travel faster than electrons, a finding that the BTE transport results will further confirm.

\begin{figure}[t] 
  \centering 
  \includegraphics[width=0.9\linewidth]{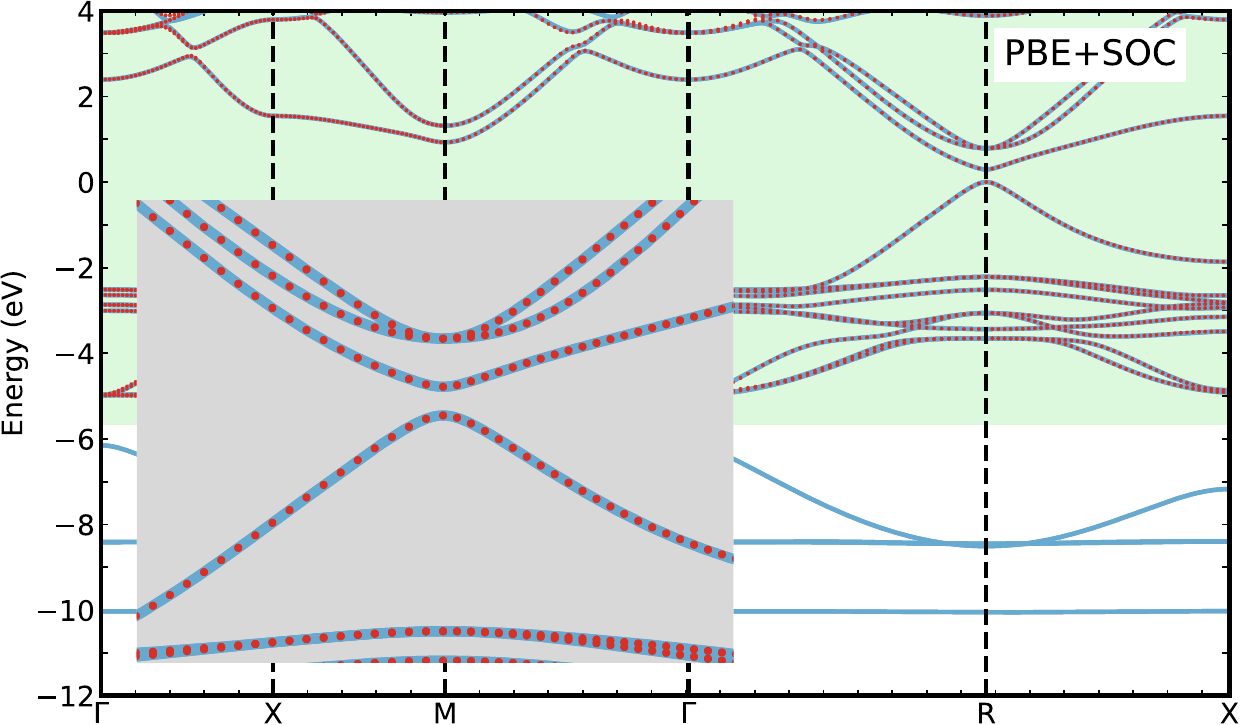} 
  \caption{\label{fig4} 
  Wannier interpolation of electronic band structures calculated using PBE + SOC. 
  The red dots represent the Wannier-interpolated band structure, where only the bands in the green-shaded area are Wannierized. 
  The gray square highlights the zoomed-in view of the band edge, which is considered for carrier transport properties.} 
\end{figure}

To reduce the computational cost of solving the BTE, we use a Wannier-Fourier interpolation~\cite{Giustino2007}. 
We use $s$ orbitals on Cs, $p$ orbitals on Sn and Br as initial guess for Wannier functions.
This corresponds to 26 and 13 Wannier functions with and without SOC, respectively. 
As shown in Fig.~\ref{fig4}, the Wannier functions accurately reproduce the band structure in the target energy window for PBE+SOC, enabling precise electron-phonon interaction calculations.
We also find similar accuracy in the case of PBE without SOC, while the case of HSE06 is not verified, given the high cost of performing direct bandstructure calculation without interpolation. 
All effective masses reported in Table~\ref{table1} are obtained from the Wannier function interpolations.

\subsection{Electron-phonon interpolation}

\begin{figure}[t] 
  \centering 
  \includegraphics[width=0.99\linewidth]{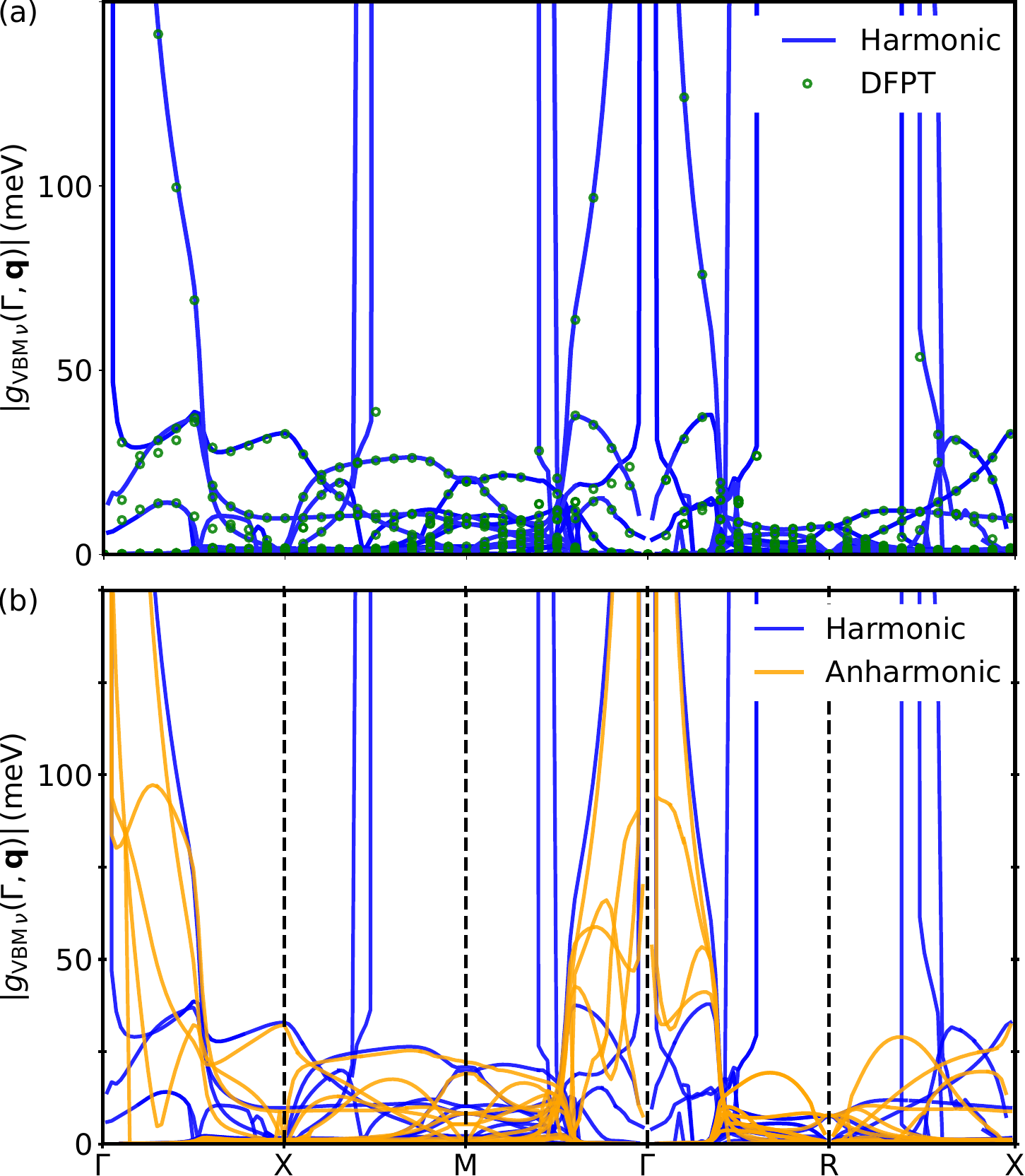} 
  \caption{\label{fig5} 
  (a) Comparison between the harmonic electron-phonon matrix elements $|g|$ computed with direct DFPT calculations (green circles) and Fourier-Wannier 
  interpolated using the \texttt{EPW} code (blue lines).
  (b) Anharmonic effect at 300~K computed with \texttt{TDEP} and Fourier-interpolated with \texttt{EPW} (orange lines) compared with harmonic interpolation (blue lines).  
  The matrix elements are shown for the valence band maximum (VBM) along the phonon momentum $\mathbf{q}$ and for all phonon branches $\nu$.
} 
\end{figure}

To assess the quality of the interpolation of the electron-phonon coupling matrix $|g_{mn\nu}(\mathbf{k}, \mathbf{q})|$, we compare in Fig.~\ref{fig5}(a) the matrix elements obtained from direct DFPT calculations with Fourier-Wannier interpolated ones along high-symmetry lines.
We find a good agreement, which confirms the quality of the Wannier interpolation.
We note that the matrix elements diverge around the $\mathbf{q}=M$ and $\mathbf{q}=R$ points due to the soft modes observed in Fig.~\ref{fig2} with harmonic calculations.
Moreover, since CsSnBr$_3$ is a polar material, electrons interact with longitudinal optical (LO) modes, resulting in a Fr\"ohlich divergence of the $|g_{mn\nu}(\mathbf{k}, \mathbf{q})|$  as \( |\mathbf{q}| \rightarrow 0 \).
Previous studies have emphasized the importance of quadrupoles in $g_{mn\nu}(\mathbf{k}, \mathbf{q})$ calculations~\cite{Brunin2020,Ponce2023}. 
Dynamical quadrupoles are nonzero in all noncentrosymmetric crystals. 
However, they can also be nonzero in centrosymmetric crystals if one or more atoms are positioned at noncentrosymmetric sites~\cite{Jhalani2020}. 
Since CsSnBr$_3$ possesses spatial inversion symmetry, quadrupoles are zero by symmetry and do not contribute, which explains the good agreement in Fig.~\ref{fig5}(a) without long-range quadrupolar corrections.

We then used the effective interatomic force constants computed with \texttt{TDEP} at 300~K and interpolate them using Wannier-Fourier interpolation, 
shown in Fig.~\ref{fig5}(b) with orange lines. 
By comparing these interpolated matrix elements with the harmonic one (blue lines), we find finite values around the $\mathbf{q}=M$ and $\mathbf{q}=R$ points,
but also a significant increase around the center of the zone for all but the Fr\"ohlich longitudinal optical mode. 
This general increase of $|g_{mn\nu}(\mathbf{k}, \mathbf{q})|$ will result in an increase in the scattering rate and a decrease in mobility.

\subsection{Carrier transport}

Finally, we compute the drift and Hall carrier transport properties of the lead-free inorganic perovskite  CsSnBr$_3$ by solving the BTE, starting from a 6 $\times$ 6 $\times$ 6 coarse $\bk$- and $\bq$-grids and interpolating the Hamiltonian, dynamical matrix, and electron-phonon matrix elements on a fine 80 $\times$ 80 $\times$ 80 $\bk$- and $\bq$-grids

\begin{figure}[t] 
  \centering 
  \includegraphics[width=0.99\linewidth]{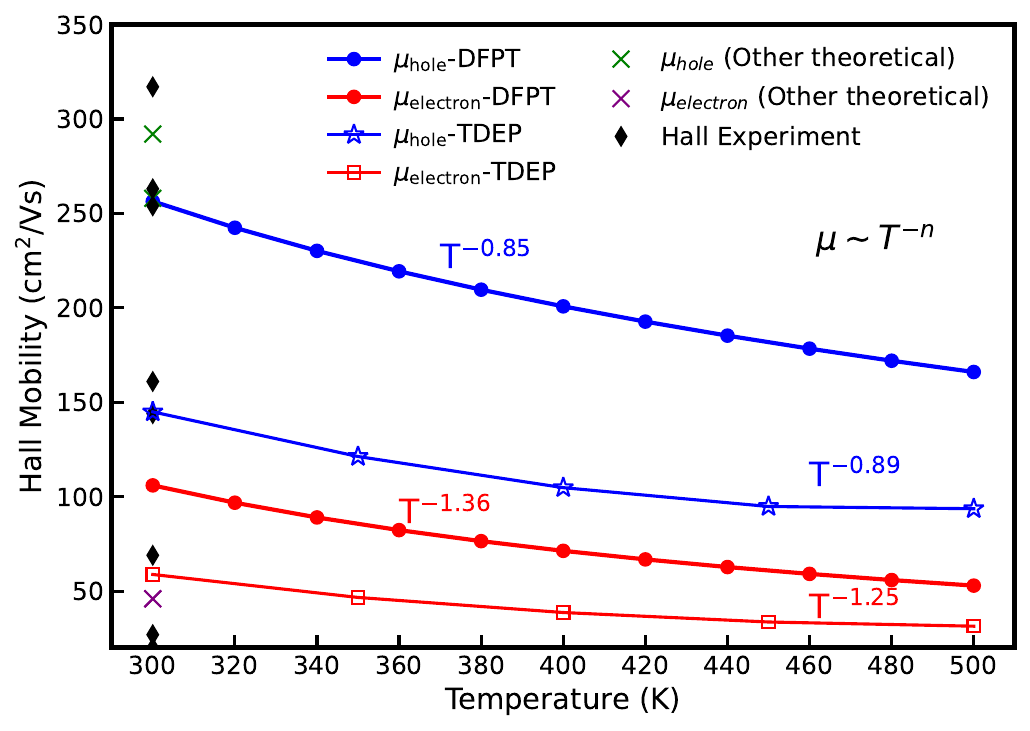} 
  \caption{\label{fig6} 
  Hall mobility calculated using an electronic band structure at the HSE06 level with SOC.
  The blue (hole) and red (electron) dots and lines represent results from harmonic DFPT-calculated phonon dispersions, while the blue stars (hole) and red squares (electron) correspond to anharmonic results from TDEP-calculated phonon dispersions.
  The temperature exponent of the Hall mobility is also reported.
  Experimental Hall data are from Ref.~\onlinecite{Jia2022} and theoretical data from Refs.~\onlinecite{Huang2013,Cao2022,Su2022,Zeng2024}.
}  
\end{figure}

Figure~\ref{fig6} presents the temperature-dependent carrier mobility of CsSnBr$_3$, calculated using the iterative Hall BTE approach, which includes the Lorenz force due to a vanishing magnetic field~\cite{Ponce2021}. 
Since the cubic phase of CsSnBr$_3$ remains stable above 292~K, we focus on mobility data within the 300~K to 500~K range.
The accuracy of carrier mobility calculations is highly sensitive to the precise description of the edges of the electronic band. 
As discussed previously, both PBE calculations with and without SOC severely underestimate the band gap, resulting in a pronounced coupling between the VBM and CBM. 
This is evidenced by the non-parabolic nature of the band edge in the PBE results. 
However, the HSE06 calculations provide a more accurate representation of the band edge, preserving its parabolic nature.
Consequently, our carrier mobility calculations are based exclusively on HSE06-calculated electronic structures with SOC. 
The Hall hole mobility consistently exceeds the electron mobility in both harmonic and anharmonic calculations.
Moreover, the addition of anharmonic effects yields lower electron and hole mobilities, but the temperature dependence is only weakly affected. 
These findings stem from the fact that harmonic phonons have soft modes that are neglected when solving the BTE, leading to an overestimation of the carrier mobilities.
We computed that the hall electron and hole mobilities at room temperature are 106 and 256~cm$^2$/Vs, respectively.
When including anharmonicity, these values decrease to 59 and 145~cm$^2$/Vs, respectively. 
We also calculated drift mobility (see Table.~S1 of the SI~\cite{Yin2025}) and found slightly lower values, indicating a temperature-dependent Hall factor ranging from 1.16 to 1.19 for holes and from 1.1 to 1.13 for electrons.

\begin{figure}[t] 
  \centering 
  \includegraphics[width=0.99\linewidth]{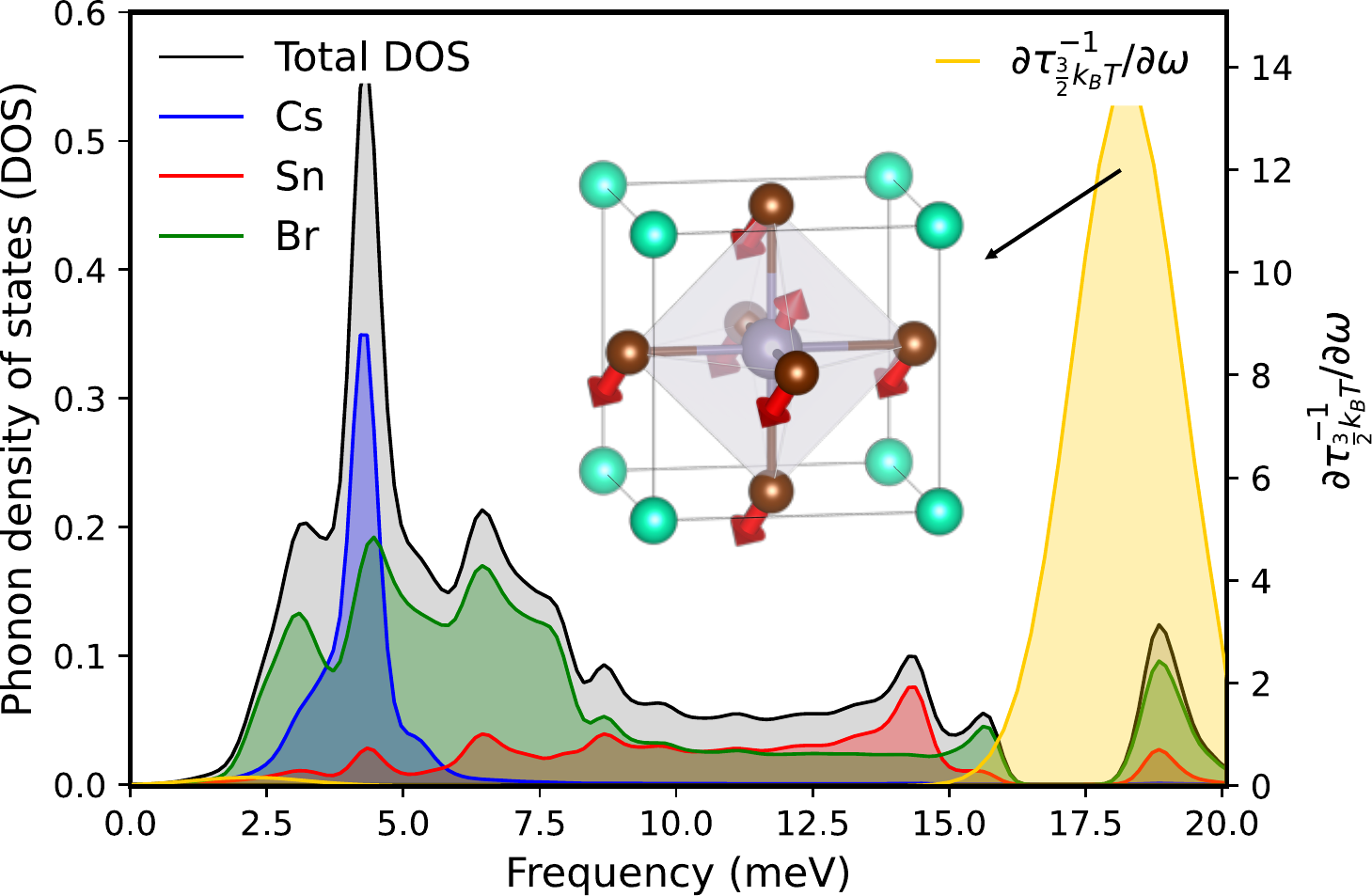} 
  \caption{Projected phonon density states (grey, blue, red, and green), hole spectral decomposition of the scattering rates (orange), and displacement vectors for the phonon mode associated with the peak located around 18~meV. 
  } 
  \label{fig7} 
\end{figure}

To understand the microscopic origin of the mobility limitation in CsSnBr$_3$, we show in Fig.~\ref{fig7} the spectral decomposition of the anharmonic result
and find that the highest LO mode exhibits the strongest electron-phonon coupling, making the most significant contribution to electron-phonon scattering processes. 
We conclude that the relatively low carrier mobility is due to the electron-phonon scatterings from the LO mode, which is related to the Pb-Br stretching modes.

Finally, we note that reports on experimental characterizations of Hall carrier mobilities of CsSnBr$_3$ are rare and, as shown in Fig.~\ref{fig6}, the mobilities of CsSnBr$_3$ vary significantly between samples and are difficult to compare to our idealized calculations. 
Still, we remark that we do fall in the experimental range.
By fitting the carrier mobility against T$^{-n}$ according to the T$^{n}$ power law in the range of 300~K - 500~K, we obtain n = 0.89 for hole and 1.25 for electron for the TDEP calculations, while we extract n = 0.85 for the hole and 1.36 for DFPT calculations. 
Overall, our transport calculations demonstrate that neglecting anharmonic effects can result in substantial errors in the prediction of carrier mobility and, more broadly, in phonon-related thermoelectric performance. 
These discrepancies are especially pronounced in materials where dynamic disorder and strong anharmonic interactions play a significant role.
This highlights the critical need to move beyond harmonic approximations when modeling materials with pronounced anharmonicity.
Incorporating fully anharmonic lattice dynamics, and even further with temperature-dependent electronic structures, will be essential for achieving reliable and predictive transport models.
The data that support the findings of this article are openly available~\cite{MCA2025}.

\section{Conclusion}
In conclusion, we developed a workflow to explore electron-phonon interactions and carrier transport in anharmonic materials. 
Our analysis of CsSnBr$_3$ reveals that its carrier mobility is predominantly constrained by electron-phonon scattering from the LO phonon mode. 
Compared to harmonic mobility calculations, the inclusion of anharmonic effects results in a reduction in the carrier mobility of 44\% for the hole and 43\% for the electron due to the enhanced electron-phonon coupling. 
We also find a weakly temperature-dependent Hall factor of around 1.1 indicating that Hall and time-of-flight experiments should yield similar results for this material. 
This work provides a valuable tool for future investigations into the transport properties of perovskites and anharmonic systems.

\begin{acknowledgments}
S. P. acknowledges support from the Fonds de la Recherche Scientifique de Belgique (FRS-FNRS). 
This work was supported by the Fonds de la Recherche Scientifique - FNRS under Grants number T.0183.23 (PDR) and T.W011.23 (PDR-WEAVE).
This publication was supported by the Walloon Region in the strategic axe FRFS-WEL-T.
Computational resources have been provided by the PRACE award granting access to MareNostrum4 at Barcelona Supercomputing Center (BSC), Spain, and Discoverer in SofiaTech, Bulgaria (OptoSpin project id. 2020225411),
by the EuroHPC JU award granting access to MareNostrum5 at Barcelona Supercomputing Center (BSC), Spain (Project ID: EHPC-EXT-2023E02-050), and by the Consortium des Équipements de Calcul Intensif (CÉCI), funded by the FRS-FNRS under Grant No. 2.5020.11,
by the Tier-1 supercomputer of the Walloon Region (Lucia) with infrastructure funded by the Walloon Region under the grant agreement n°1910247.
\end{acknowledgments}

\end{document}